\documentclass[sigconf,numbers,sort&compress]{acmart}
\usepackage{CJKutf8}
\usepackage{CJK}
\usepackage{multirow}
\usepackage{siunitx}
\usepackage{adjustbox}
\newcommand{\kaiti}[1]{\begin{CJK*}{UTF8}{gkai} #1 \end{CJK*}}
\usepackage[utf8]{inputenc}
\usepackage{makecell}
\usepackage{booktabs} 
\usepackage{graphicx}
\usepackage{amsmath}
\usepackage{caption}
\usepackage{subcaption}
\usepackage{multirow}
\AtBeginDocument{%
  }

\copyrightyear{2025}
\acmYear{2025}
\setcopyright{acmlicensed}\acmConference[CIKM '25]{Proceedings of the 34th
ACM International Conference on Information and Knowledge
Management}{November 10--14, 2025}{Seoul, Republic of Korea}
\acmBooktitle{Proceedings of the 34th ACM International Conference on
Information and Knowledge Management (CIKM '25), November 10--14, 2025,
Seoul, Republic of Korea}
\acmDOI{10.1145/3746252.3760880}
\acmISBN{979-8-4007-2040-6/2025/11}

\begin{document}

\title{H-PRM: A Pluggable \underline{H}otword \underline{P}re-\underline{R}etrieval \underline{M}odule for Various Speech Recognition Systems}

\author{Huangyu Dai}
\authornote{Equal Contribution.}
\affiliation{
  \institution{Kuaishou Technology}
  \city{Hangzhou}
  \country{China}
}
\email{dai244678786@gmail.com}

\author{Lingtao Mao}
\authornotemark[1]
\authornote{Corresponding author.}
\affiliation{
  \institution{Kuaishou Technology}
  \city{Hangzhou}
  \country{China}
}
\email{mltzju@163.com}

\author{Ben Chen}
\authornotemark[1]
\affiliation{%
  \institution{Kuaishou Technology}
  \city{Hangzhou}
  \country{China}
}
\email{benchen4395@gmail.com}

\author{Zihan Wang}
\affiliation{
  \institution{Kuaishou Technology}
  \city{Hangzhou}
  \country{China}
}
\email{2101213096@stu.pku.edu.cn}

\author{Zihan Liang}
\affiliation{
  \institution{Kuaishou Technology}
  \city{Hangzhou}
  \country{China}
}
\email{liangzih@seas.upenn.edu}

\author{Ying Han}
\affiliation{
  \institution{Zhejiang Gongshang University}
  \city{Hangzhou}
  \country{China}
}
\email{hanyinglucy@mail.zjgsu.edu.cn}

\author{Chenyi Lei}
\affiliation{
  \institution{Kuaishou Technology}
  \city{Hangzhou}
  \country{China}
}
\email{leichy@mail.ustc.edu.cn}

\author{Han Li}
\affiliation{
  \institution{Kuaishou Technology}
  \city{Beijing}
  \country{China}
}
\email{lihan08@kuaishou.com}

\renewcommand{\shortauthors}{Dai et al.}

\begin{abstract}
Hotword customization is crucial in ASR to enhance the accuracy of domain-specific terms. It has been primarily driven by the advancements in traditional models and Audio large language models (LLMs). However, existing models often struggle with large-scale hotwords, as the recognition rate drops dramatically with the number of hotwords increasing. In this paper, we introduce a novel hotword customization system that utilizes a hotword pre-retrieval module (H-PRM) to identify the most relevant hotword candidate by measuring the acoustic similarity between the hotwords and the speech segment. This plug-and-play solution can be easily integrated into traditional models such as SeACo-Paraformer, significantly enhancing hotwords post-recall rate (PRR). Additionally, we incorporate H-PRM into Audio LLMs through a prompt-based approach, enabling seamless customization of hotwords. Extensive testing validates that H-PRM can outperform existing methods, showing a new direction for hotword customization in ASR.
\end{abstract}



\maketitle

\section{Introduction}

End-to-end ASR models have significantly improved speech recognition accuracy in the past decade~\cite{graves2012sequence,chorowski2015attention,graves2013speech,chan2016listen,vaswani2017attention,gulati2020conformer,tian2020spike,fan2021cass,fan2021improved,deng2022improving,gao2022paraformer, radford2023robust, chu2024qwen2}.
However, advanced ASR models still face challenges in accurately recognizing hotwords (e.g., people's names, place names), due to their infrequent appearance in training data. To address this, effective hotword customization techniques have been developed~\cite{han2021cif,huang2023contextualized,yang2023two}, using shallow~\cite{williams2018contextual,zhao2019shallow,gourav2021personalization} and deep~\cite{han2022improving,munkhdalai2022fast,sainath2023improving,pundak2018deep,shi2024seaco} fusion methods to integrate hotword list into ASR models. 
With the rise of Audio LLMs, new approaches have emerged for hotword customization, such as the KWS-Whisper system~\cite{li2024multitask} and CTC-Assisted LLM-Based Contextual ASR~\cite{yang2024ctc}, which use retrieval modules and prompts to enhance ASR results by effectively integrating hotwords.

However, as the number of hotwords increases, mutual interference among them increases the overall mixed error rate (MER) for both fusion and retrieval-based approaches. To address this challenge, we propose the hotword pre-retrieval module (H-PRM). This module estimates the occurrence probability of each hotword in the input speech by combining the ASR result with hotword texts through phonemic embeddings and efficiently retrieves top-N candidate hotwords for seamless integration into contextual ASR systems or Audio large language models (LLMs).

\begin{figure*}[htbp]
    \centering
    \includegraphics[width=\textwidth]{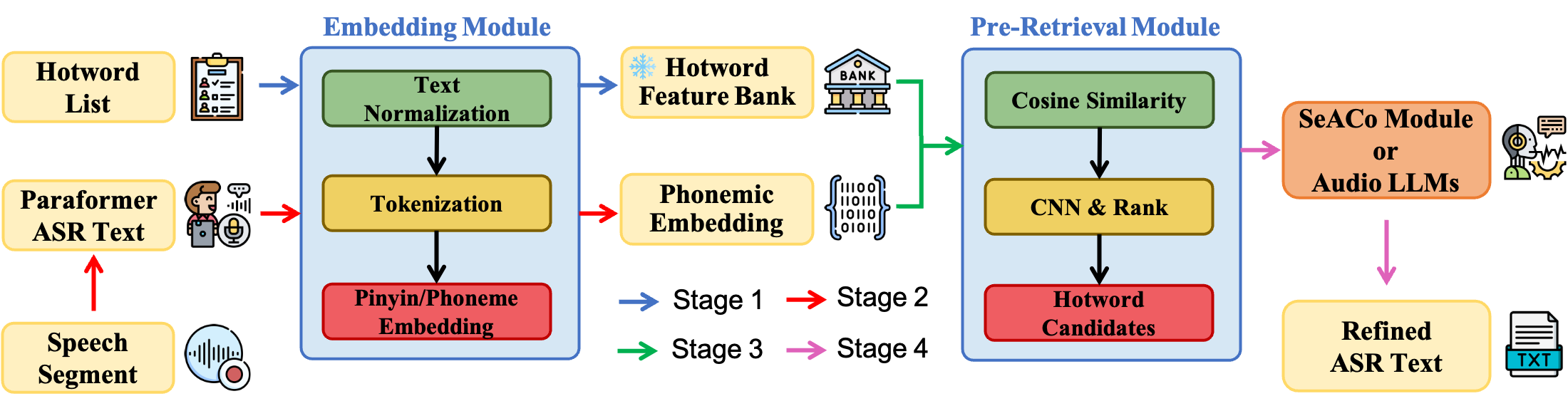}
    \caption{Overview of the proposed hotword customization system, including Paraformer ASR Model, Embedding Module, Hotword Pre-Retrieval Module, and SeACo Module (or Audio LLMs). }
    \label{fig:MainFlows}
\vspace{-0.5em}
\end{figure*}

Extensive experimental results on both public (Aishell-dev/test) and self-constructed datasets (Common-voice-zh/en) have demonstrated the effectiveness of H-PRM, significantly improving the hotword post-recall rate (PRR) and reducing the MER. For example, on the Common-voice-zh dataset, our module improves PRR by 29.07\% and reduces MER by 30.54\% in SeACo-Paraformer. Additionally, introducing H-PRM enables the Whisper-small and Qwen2-Audio-Instruct to achieve an average MER reduction of 29.44\% across four datasets. These results demonstrate the effectiveness of our approach in enhancing hotword recognition accuracy across diverse ASR models, even with extensive hotword lists. Corresponding datasets and codes will be made publicly available after publication.


\vspace{-0.5em}
\begin{figure}[htbp]
    \centering
    \begin{subfigure}[b]{0.46\textwidth}
        \centering
        \includegraphics[width=\textwidth]{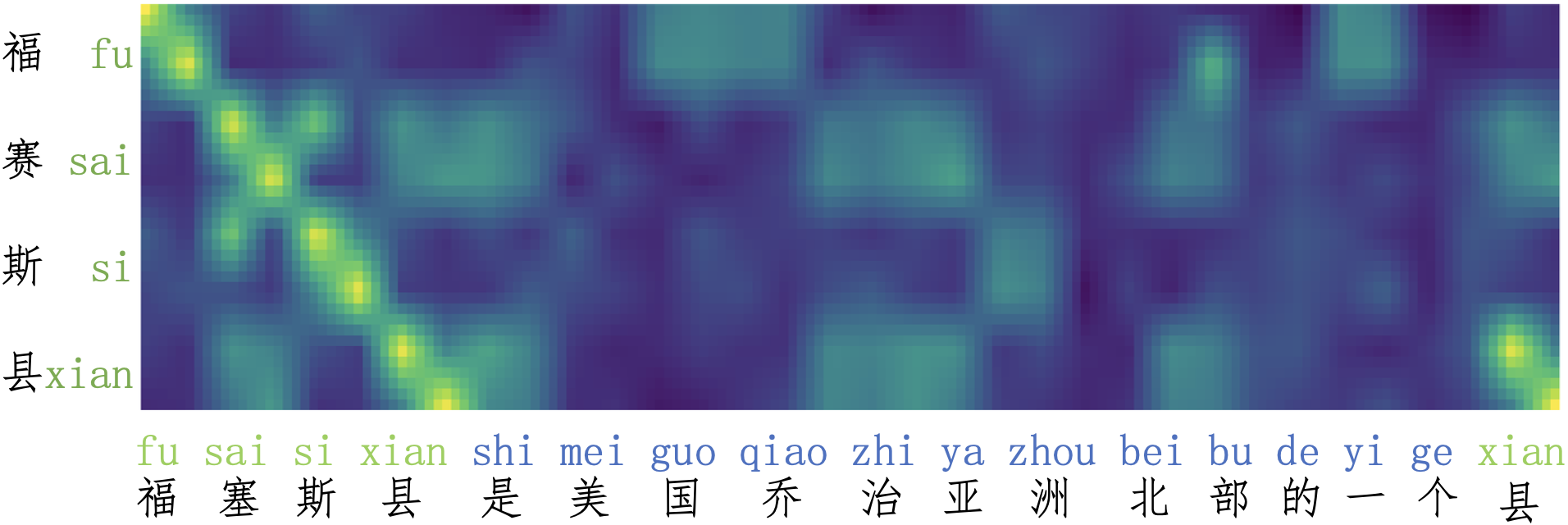}
        \caption{}
    \end{subfigure}
    \hfill
    \begin{subfigure}[b]{0.46\textwidth}
        \centering
        \includegraphics[width=\textwidth]{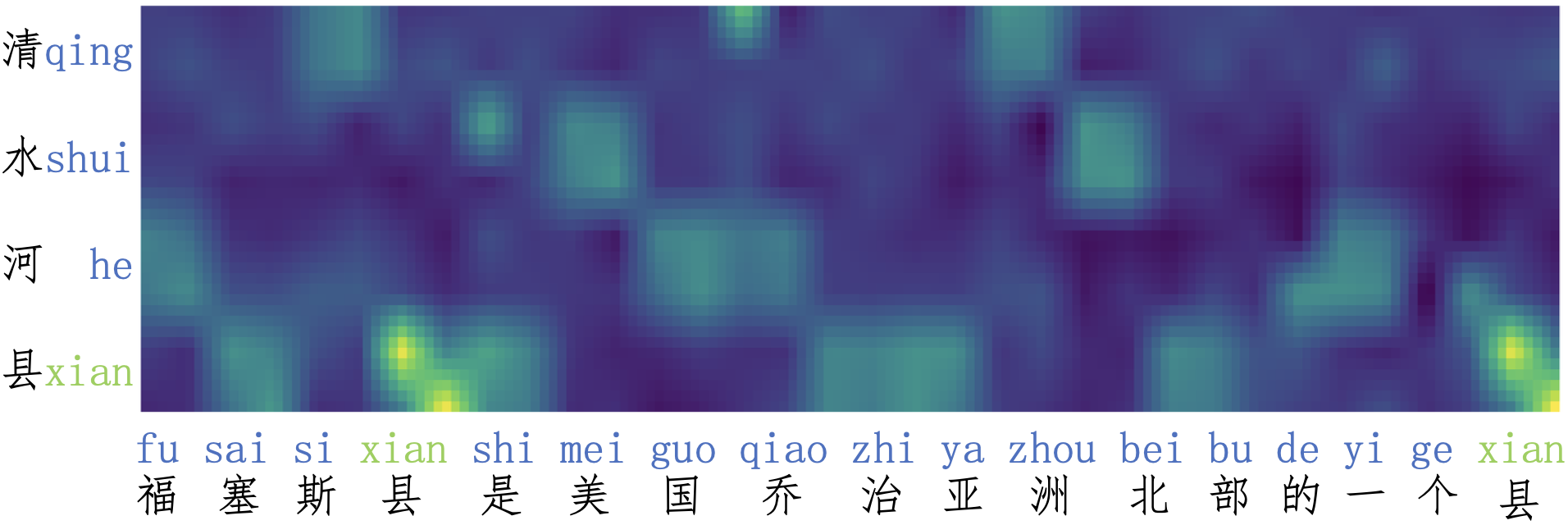}
        \caption{}
    \end{subfigure}
    \caption{Cosine similarity heatmaps of Pinyin embedding between hotword and ASR result. (a) ground truth hotword. (b) unrelated hotword.}
    \label{fig:heatmaps}
\vspace{-2em}
\end{figure}

\section{Related Work}
\subsection{SeACo-Paraformer}
SeACo-Paraformer is an advanced method for customizing hotwords in NAR ASR systems such as the Paraformer. It enables hotword customization during inference by using bias encoding and attention mechanisms. This method embeds hotwords with a bias encoder, integrates these embeddings into the acoustic and decoder hidden states via multi-head attention, and generates biased probabilities to combine with Paraformer outputs. To handle large-scale hotword lists, SeACo-Paraformer employs an ASF~\cite{shi2024seaco} strategy, filtering hotwords based on attention scores for effective and accurate activation even with extensive lists.

\subsection{Audio LLMs}
Audio LLMs, such as Whisper and Qwen2-Audio, have significantly advanced audio processing performance. Whisper inserts prompts at the start of audio tokens to ensure consistent influence, while Qwen2-Audio-Instruct dynamically adjusts the insertion of prompts using special tokens tailored to specific task requirements. Both models use special tokens to facilitate seamless task switching, allowing for transcription, translation, and audio language identification. Besides, they also introduce a new paradigm for hotword customization using customized prompts.

\section{Methodology}

\subsection{System overview}


The proposed hotword customization system functions in four stages (see Figure~\ref{fig:MainFlows}). 
In the first stage, normalization and tokenization convert the hotword list into Pinyin or phonemes, with the corresponding phonemic embeddings stored in the hotword feature bank. These operations are standard in the frontend module of Text-To-Speech models, and our system utilizes the SAM-BERT~\cite{li2020robutrans} frontend module. In the second stage, the Paraformer model is employed to generate a recognition result for the speech segment. This recognition result is then converted into a phonemic embedding, similar to the hotwords. In the third stage, the H-PRM module selects the top-N hotwords candidates by processing the similarity matrix between the phonemic embeddings of ASR text and hotwords. Finally, hotword customization methods produce refined ASR text using both the hotword candidates and speech information as input. Specifically, we use the Paraformer output embedding for the SeACo module and the original speech for Audio LLMs.

\begin{table}[htbp]
\caption{Prompts for the Whisper-small and Qwen2-Audio-Instruct models.}
\centering
\renewcommand{\arraystretch}{1.1} 
\begin{tabular}{>{\raggedright}p{2cm}|p{5.2cm}}
\toprule
\textbf{Whisper\\-small \\Prompt} & \kaiti{今天演讲的主题是这个呃，实体1、实体2、实体3。好，那我就继续讲。} \\
\hline
\textbf{Qwen2\\-Audio\\-Instruct\\ Prompt} & \kaiti{Audio1 \texttt{<|BOS|><|AUDIO|>}\texttt{<|EOS|>}请对上述音频进行中文语音识别，重点关注热词列表[实体1, 实体2, 实体3]，没有发现热词请直接输出完整的音频语音识别结果。请严格按照以下格式输出：\{“音频内容”: 语音识别结果\}} \\
\bottomrule
\end{tabular}
\label{tab:Audio LLMs prompt}
\vspace{-1em}
\end{table}

\begin{table*}[tbp]
\caption{\textbf{MER (\%)} and \textbf{PRR (\%)} of different hotword customization methods (HCM) and ASR models with top-N hotwords.}
\vspace{-0.3em}
\centering
\begin{tabular}{>{\raggedright}
 p{3.4cm}|p{1.8cm}
|p{1.1cm}|p{1.1cm}
|p{1.1cm}|p{1.1cm}
|p{1.1cm}|p{1.1cm}
|p{1.1cm}|p{1.1cm}}
\toprule
\multicolumn{2}{c|}{\textbf{Methods}} & \multicolumn{2}{c|} {\textbf{Aishell-dev}} & \multicolumn{2}{c|} {\textbf{Aishell-test}} & \multicolumn{2}{c|} {\textbf{Common-voice-zh}} & \multicolumn{2}{c} {\textbf{Common-voice-en}} \\

\cmidrule(lr){1-2}
\cmidrule(lr){3-4}
\cmidrule(lr){5-6}
\cmidrule(lr){7-8}
\cmidrule(lr){9-10}

\textbf{ASR Models} & \textbf{HCM} 
& \textbf{MER} & \textbf{PRR}  
& \textbf{MER} & \textbf{PRR} 
& \textbf{MER} & \textbf{PRR} 
& \textbf{MER} & \textbf{PRR} \\
\midrule
\hline

\multirow{3}{*}{\textbf{Paraformer}} & \textbf{-} 
& 5.39 & 56.12 & 5.28 & 58.82 & 9.44 & 70.41 & - & - \\

\
 & \textbf{SeACo~\textsuperscript{\cite{shi2024seaco}}} 
& 1.98 & 94.90 & 2.34 & 92.50 & 9.45 & 70.63 & - & - \\

 & \textbf{H-PRM} 
& \textbf{1.77} & \textbf{95.88} & \textbf{2.16} & \textbf{95.11} & \textbf{6.57} & \textbf{93.90} & - & - \\
\hline

\multirow{3}{*}{\textbf{Whisper-small}} & \textbf{-} 
& 16.24 & 36.84 & 18.52 & 41.12 & 21.20 & 51.81 & 9.78 & 60.02 \\

 & \textbf{OV-KWS~\textsuperscript{\cite{li2024multitask}}} 
& 14.54 & 51.23 & 15.51 & 53.42 & 19.78 & 67.22 & 8.86 & 62.89 \\

 & \textbf{H-PRM} 
& \textbf{9.88} & \textbf{86.98} & \textbf{11.10} & \textbf{89.99} & \textbf{16.40} & \textbf{87.13} & \textbf{7.60} & \textbf{68.90} \\
\hline

\multirow{2}{*}{\textbf{Qwen2-Audio-Instruct}} & \textbf{-} 
& 6.44 & 69.46 & 6.60  & 68.89 & 7.53 & 75.12 & 6.53 & 59.42 \\
 & \textbf{H-PRM} 
& \textbf{3.24} & \textbf{88.94} & \textbf{4.66} & \textbf{87.84} & \textbf{6.02} & \textbf{85.61} & \textbf{5.73} & \textbf{67.80} \\
\bottomrule
\end{tabular}
\label{tab:main_experiment}
\vspace{-0.6em}
\end{table*}

\begin{table}[tbp]
\centering
\caption{Test sets for general and customization ASR}
\vspace{-0.3em}
\begin{tabular}{lcc}
\toprule
\textbf{Dataset} & \textbf{Utterances} & \textbf{Hotwords} \\
\midrule
Aishell--dev & 1319  & 591 \\
Aishell--test & 781  & 394 \\
Common-voice-zh (Ours) & 4090  & 3800 \\
Common-voice-en (Ours) & 794 & 565 \\
\bottomrule
\end{tabular}
\label{table:dataset info}
\vspace{-0.5em}
\end{table}


\subsection{H-PRM design}


The H-PRM employs a binary classification convolutional neural network (CNN) to assess the similarity between the continuous pronunciations of each hotword and the input speech segment and selects the top-N hotword candidates with the highest similarity.
Specifically, the phonemic embeddings of hotwords are sequentially extracted from a predefined hotword feature bank. For each hotword, a cosine similarity matrix is computed between its phonemic embedding and the phonemic embedding obtained from the ASR result. CNN then processes this similarity matrix to produce a normalized similarity score. Finally, all hotwords are ranked according to their similarity scores, and the top-N highest-scoring matches are selected as the module output. When a hotword is phonetically similar to the part of the speech segment, the similarity matrix displays a distinct diagonal bright line, and CNN tends to output a high score correspondingly, as illustrated in Figure~\ref{fig:heatmaps} (a). In contrast, if no hotword is phonetically similar to the speech segment, the similarity matrix appears disordered, as shown in Figure~\ref{fig:heatmaps} (b).

\subsection{Training trick for the H-PRM}

To enhance the binary classification performance of the CNN model for the H-PRM module, we propose an iterative hard-sample mining strategy that progressively refines the training data through three key stages. First, we construct initial positive samples by pairing ground-truth hotwords with ASR-recognized texts that exhibit a MER between 2\% and 20\%, simulating common speech recognition errors. Second, after each training iteration, we dynamically generate challenging negative samples through retrieval-based analysis. If the hotword retrieved from the top-1 mismatches the ground truth, it forms a negative pair with the input text; otherwise, the second-ranked incorrect retrieval is selected. Finally, in the last iteration, we further improve model robustness to pronunciation variations by creating augmented positive samples through controlled character-level modifications of ground-truth hotwords, while pairing them with their ground-truth ASR text. This multi-stage approach systematically enhances the discriminative power of the CNN by exposing it to increasingly sophisticated edge cases while maintaining acoustic relevance.



\begin{figure}[htbp]
    \centering
    \begin{subfigure}[t]{0.222\textwidth}
        \centering
        \includegraphics[width=\textwidth]{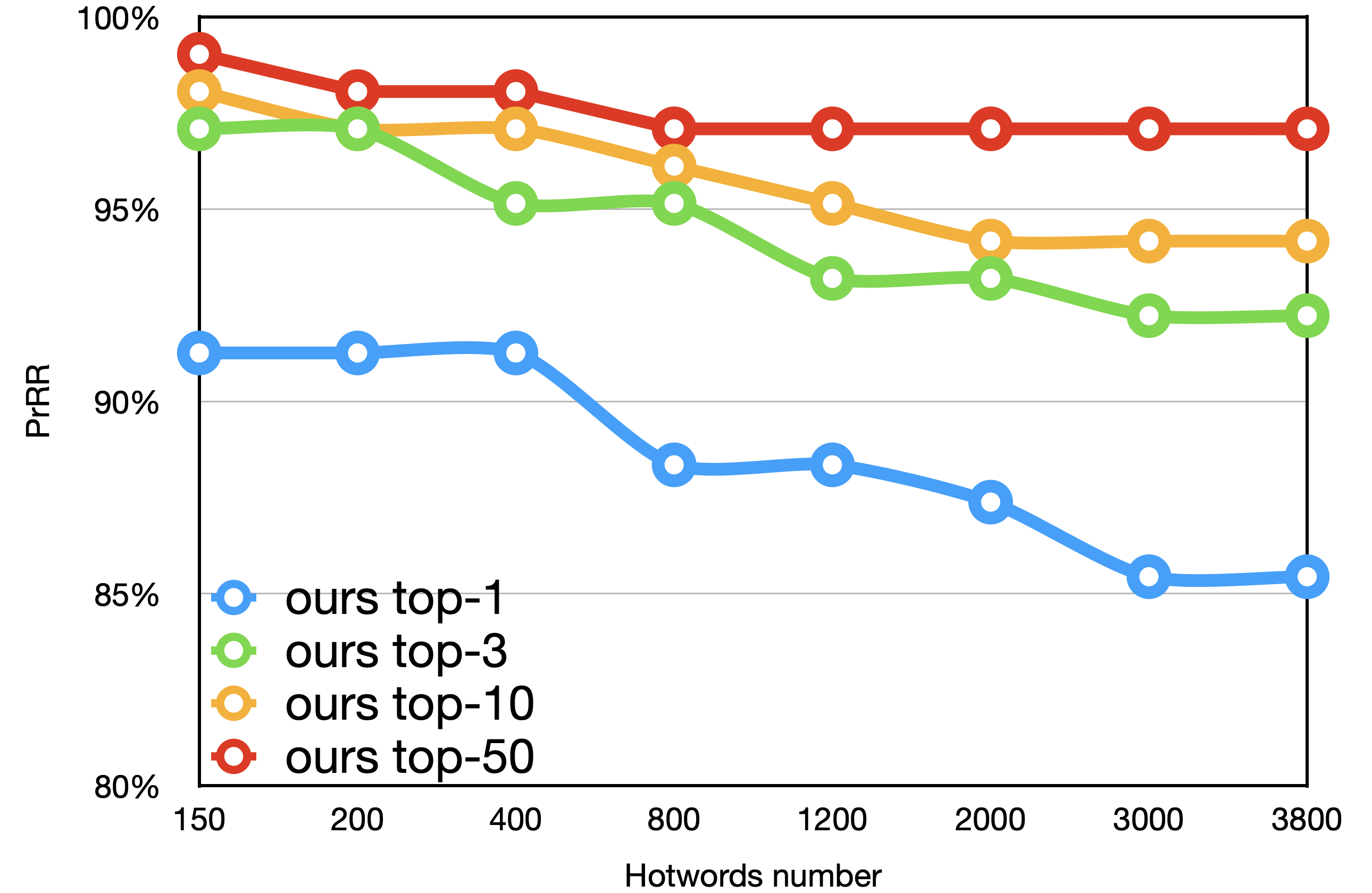}
        \caption{}
        \label{fig:subfig1}
    \end{subfigure}
    \hfill
    \begin{subfigure}[t]{0.222\textwidth}
        \centering
        \includegraphics[width=\textwidth]{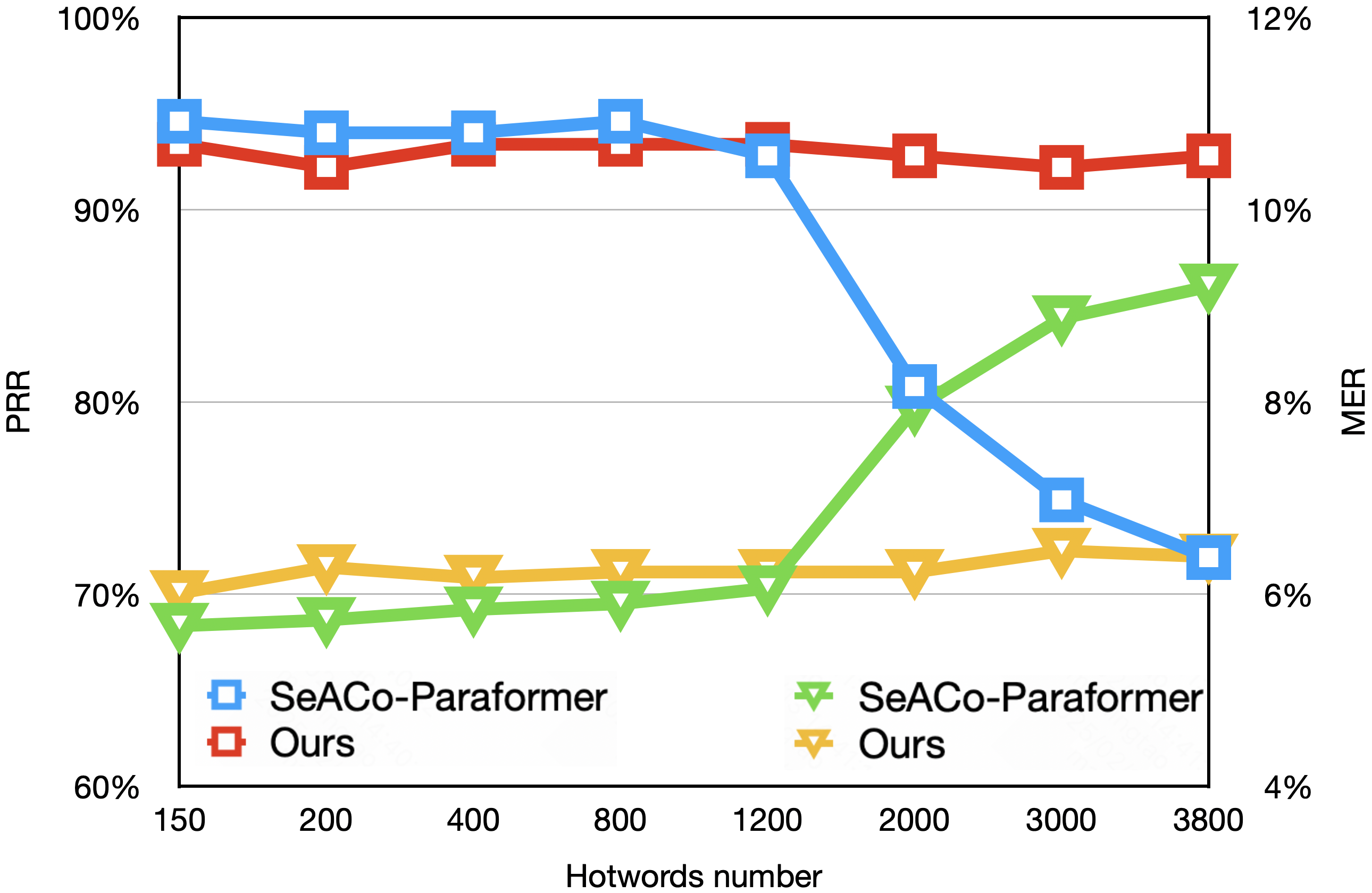}
        \caption{}
        \label{fig:subfig2}
    \end{subfigure}
    \caption{Trends of different metrics with varying numbers of hotwords: (a) \textbf{PrRR (\%)}. (b) \textbf{PRR (\%)} and \textbf{MER (\%)}.}
    \label{fig:different hotwords numbers}
    \vspace{-1.2em}
\end{figure}


\begin{table}[htbp]
    \centering
    \vspace{-0.2em}
    \caption{The performance varies across different models with the top-N recall hotwords changing,  measured by \textbf{PrRR (\%) / PRR (\%) / PF1 (\%) / MER (\%)}.}
    \begin{adjustbox}{max width=0.95\textwidth/2}
    \begin{tabular}{ccccc}
        \toprule
        \textbf{Model} & \textbf{Hotwords} & \textbf{Aishell-dev} & \textbf{Common-voice-zh} \\
        \midrule
        \multirow{5}{*}{\makecell{SeACo-\\Paraformer}} & $\times$ 
        & - / 56.1 / 71.8 /5.39 & - / 70.4 / 81.9 / 9.44\\
        & GT & \underline{100} / \underline{98.0} / \underline{99.0} / \underline{1.48} & \underline{100} / \underline{96.5} / \underline{97.8} / \underline{5.93} \\
        & Top-1  & 92.6 / 90.8 / 94.8 / 2.31 & 87.4 / 86.2 / 90.8 / 7.64 \\
        & Top-3  & 96.4 / 95.1 / 97.0 / 1.90 & 94.6 / 91.6 / 93.5 / 6.97 \\
        & Top-10 & 97.9 / 95.8 / \textbf{97.4} / 1.79 & 96.4 / 92.9 / 94.0 / 6.82 \\
        & Top-50 & \textbf{99.3} / \textbf{95.9} / \textbf{97.4} / \textbf{1.77} & \textbf{98.2} / \textbf{93.9} / \textbf{94.9} / \textbf{6.57} \\
        \midrule
        \multirow{5}{*}{\makecell{Whisper-\\small}} 
        & $\times$ & - / 36.8 / 53.7 / 16.24 & - / 83.3 / 88.4 / 18.82 \\
        & GT &  \underline{100} /  \underline{90.8} /  \underline{95.1} /  \underline{9.08} &  \underline{100} /  \underline{92.6} /  \underline{95.6} /  \underline{14.09} \\
        & Top-1 & 92.6 / 83.8 / 90.9 / 9.98 & 87.4 / 79.2 / 87.3 / \textbf{16.35} \\
        & Top-3 & 96.4 / \textbf{87.0} / \textbf{91.8} / \textbf{9.88} & 94.6 / \textbf{87.1} / \textbf{90.6} / 16.40 \\
        & Top-10 & 97.9 / 84.3 / 90.2 / 10.90 & 96.4 / 86.2 / 89.7 / 16.46 \\
        & Top-50 & \textbf{99.3} / 38.8 / 55.4 / 15.12 & \textbf{98.2} / 53.3 / 68.1 / 20.79 \\
        \midrule
        \multirow{5}{*}{\makecell{Qwen2-Audio\\-Instruct}} & $\times$ 
        & - / 71.3 / 83.1 / 5.16 & - / 75.1 / 85.0 / 7.53 \\
        & GT & \underline{100} / \underline{93.5} / \underline{96.5} / \underline{2.68} & \underline{100} / \underline{91.2} / \underline{94.8} / \underline{5.19} \\
        & Top-1 & 92.6 / 88.9 / \textbf{93.8} / \textbf{3.24} & 87.4 / 85.6 / 91.4 / \textbf{6.02} \\
        & Top-3 & 96.4 / \textbf{90.6} / 93.2 / 3.76 & 94.6 / 88.2 / \textbf{91.7} / 6.42 \\
        & Top-10 & 97.9 / 90.1 / 84.2 / 8.15 & 96.4 / 88.8 / 86.9 / 9.36 \\
        & Top-50 & \textbf{99.3} / 90.4 / 56.0 / 29.42 & \textbf{98.2} / \textbf{89.9} / 65.2 / 25.88 \\
        \bottomrule
    \end{tabular}
    \end{adjustbox}
    \label{tab:topn wer}
    \vspace{-1em}
\end{table}

\begin{table*}[tbp]
\caption{\textbf{PrRR (\%)} of H-PRM using different modal embeddings (a for audio, t for text, and p for phoneme).}
\vspace{-0.6em}
\centering
\begin{tabular}{>{\raggedright}
 p{3.5cm}|p{1.5cm}
|p{0.9cm}|p{0.9cm}|p{0.9cm}|p{0.9cm}
|p{0.9cm}|p{0.9cm}|p{0.9cm}|p{0.9cm}}
\toprule
\multicolumn{2}{c|}{\textbf{Embeddings}} & \multicolumn{4}{c|} {\textbf{Aishell-dev}} & \multicolumn{4}{c} {\textbf{Common-voice-zh}} \\
\cmidrule(lr){1-2}
\cmidrule(lr){3-6}
\cmidrule(lr){7-10}
\textbf{model} & \textbf{modal} 
& \textbf{R@1}  
& \textbf{R@3} 
& \textbf{R@10} 
& \textbf{R@50} 
& \textbf{R@1} 
& \textbf{R@3} 
& \textbf{R@10} 
& \textbf{R@50} 
 \\
\midrule
\hline
\textbf{Roberta-wwm-zh} & \textbf{t->t} 
& 37.83 & 53.07 & 64.28 & 78.32  & 34.43 & 45.82 & 56.80 & 73.06\\

\textbf{KWS-Whisper-small~\textsuperscript{\cite{li2024multitask}}} & \textbf{a->a} 
& 35.64 & 49.72 & 62.19 & 74.64  & 32.57 & 42.53 & 53.89 & 70.34\\


\textbf{Qwen2-Audio-Instruct} & \textbf{a->t} 
& 0.23 & 0.68 & 2.27 & 9.48 & 0.10 & 0.22 & 0.68 & 3.25\\

\textbf{Ours (SAM-BERT)} & \textbf{p->p} 
& \textbf{92.57} & \textbf{96.36} & \textbf{97.88} & \textbf{99.32} & \textbf{87.43} & \textbf{94.57} & \textbf{96.38} & \textbf{98.17}\\
\bottomrule
\end{tabular}
\label{tab:diff_modal}
\vspace{-0.7em}
\end{table*}






\subsection{Prompt for Audio LLMs}

For Audio LLMs, hotword customization requires more than just recalling hotwords via the H-PRM. LLMs must understand the task through prompts, combining acoustic information to recognize speech segments accurately. For the Whisper, the prompt by Y. Li~\cite{li2024multitask} was employed. For the Qwen2-Audio series, Qwen2-Audio-Instruct was chosen. The prompt for Qwen2-Audio-Instruct was generated and refined by GPT-4o based on the task description. The optimized prompt is presented in Table~\ref{tab:Audio LLMs prompt}.


\section{Experiments}

\subsection{Data Introduction}

In our experiments, four datasets are utilized to evaluate the performance of our model. Among them, two are open-source datasets, Aishell-dev and Aishell-test, from which we remove 2\% of erroneous data. The remaining two are self-constructed datasets, called Common-voice-zh/en, which were collected using LLMs from the Common Voice Corpus. Specifically, we use the Common Voice Corpus 19.0~\footnote{https://commonvoice.mozilla.org/zh-CN/datasets} as the initial corpus. Hotwords are extracted using the Paraformer and Qwen2.5-32B-Instruct models. Initially, the Paraformer is employed to recognize each speech segment, discarding segments with MER greater than 10\%. Subsequently, the accurate speech recognition results are fed into the Qwen2.5-32B-Instruct model for hotword extraction. When hotwords are detected, they are extracted to construct the custom Common-voice datasets. To create the dataset for training the H-PRM, we collect data from the Common Voice Corpus. When the LLM fails to extract hotwords, random words are selected to build our training dataset. Detailed information about these datasets is provided in Table~\ref{table:dataset info}.

\subsection{Experimental Setup and Evaluation Metrics}

For efficiency consideration, we employ a lightweight five-layer CNN architecture for the H-PRM. The network comprises layers with 16, 32, 64, 128, and 128 channels, each using a kernel size of \(3 \times 3\). Two fully connected layers are incorporated following the convolutional layers, culminating in a two-class output. The model is trained using the AdamW optimizer over 50 epochs with a learning rate of \(1 \times 10^{-4}\) and a batch size of 32. 

We evaluate the performance of our H-PRM and ASR system using three primary metrics: pre-recall rate (PrRR), PRR, and MER. MER is a modified version of the character error rate (CER) that can handle code-switching between English and Chinese by treating each Chinese character and each English word as a single unit. Additionally, PrRR represents the sentence-level recall rate of hotwords by the H-PRM, while PRR measures the word-level recall rate of hotwords after speech recognition. To further assess the model's performance in hotword customization, we also use post-precision rate (PPR) and post-F1 score (PF1) as additional metrics.

\subsection{Experimental Results}

As shown in Table~\ref{tab:main_experiment}, our method consistently enhances performance across all models and datasets, as evidenced
by PRR, MER.

For the Paraformer, our method achieves an average PRR improvement of 12.27\% and a MER reduction of 16.27\% across the Aishell-dev, Aishell-test, and Common-voice-zh datasets. In scenarios with a large number of hotwords, methods like SeACo-Paraformer struggle to recognize hotwords effectively. For example, in the Common-voice-zh dataset, SeACo-Paraformer achieves a PRR of only 70.63\%, whereas our H-PRM significantly improves it to 93.90\%, reducing the MER from 9.45\% to 6.57\%. These results demonstrate that our H-PRM effectively reduces conflicts between hotwords and enhances hotword recognition performance.


For Audio-LLMs like Whisper-small and Qwen2-Audio-Instruct, our method is effective for both Chinese and English. Whisper-small's MER drops from 18.65\% to 12.46\% on Chinese datasets and from 9.78\% to 7.60\% on English datasets. However, the OV-KWS module~\cite{li2024multitask} shows limited improvement without end-to-end joint training. For Qwen2-Audio-Instruct, MER is reduced from 6.52\% to 3.95\%, and PRR improves from 69.18\% to 88.39\% on the open-source dataset Aishell. Overall, our method is plug-and-play, effectively working with ASR models for hotword customization.

To study the impact of the number of hotwords on the performance of H-PRM, we select 103 speech samples from the Common-voice-zh dataset as the entire test set, which initially includes 150 hotwords. During the experiment, we gradually enlarge the hotword list with irrelevant hotwords, expanding it to a size of 3,800. Figure~\ref{fig:different hotwords numbers} illustrates the change in PRR and MER as the number of hotwords increases. Specifically, the top-50 PrRR of our H-PRM decreases only slightly from 99.0\% to 97.0\% as the number of hotwords increases in Figure~\ref{fig:different hotwords numbers} (a), demonstrating a minimal reduction of just 2\%. This indicates its potential to manage large hotword sets effectively. Figure~\ref{fig:different hotwords numbers} (b) shows that without our module, the SeACo-Paraformer's PRR drops significantly from 94.6\% to 71.9\%, and the MER rises from 5.67\% to 9.20\%. However, with our H-PRM, PRR remains at 93\%, and the MER only slightly increases to 6.39\%. This experiment demonstrates that our method effectively supports hotword customization tasks, maintaining robust performance even with an extensive hotword list.

Furthermore, in Table~\ref{tab:topn wer}, we analyze the impact of changes in the top-N recall hotwords on PRR, PF1, and MER across various models. The results indicate that SeACo-Paraformer achieves optimal performance at top-50, with PPR of 95.9\%, PF1 of 97.4\%, and MER of 1.77\%. In contrast, the performance of Whisper-small and Qwen2-Audio-Instruct exhibits variability across different datasets. Generally, Whisper-small demonstrates optimal performance with the top-3 hotwords, whereas Qwen2-Audio-Instruct achieves superior results with only the top-1 hotword. These findings underscore the importance of selecting an appropriate number of hotwords to maximize model performance. Meanwhile, they also suggest that using prompts for hotword customization in Audio LLMs may not be the ultimate solution and warrants further investigation.

Table~\ref{tab:diff_modal} shows that the matching between phonemic embeddings (p→p) outperforms other modal embeddings in the H-PRM module, achieving R@1 rates of 92.57\% on Aishell-dev and 87.43\% on Common-voice-zh. This performance is nearly double that of audio-to-audio (a→a) and text-to-text (t→t) methods. In contrast, the audio-to-text (a→t) method exhibits a critically low performance, with R@50 values of only 9.48\% on Aishell-dev and 3.25\% on Common-voice-zh, rendering it practically unusable. These results highlight the difficulty of matching hotwords with speech segments and the challenges of cross-modal alignment. Therefore, a purely acoustic approach with phonemic embeddings proves most effective for hotword pre-retrieval at the current time.

\section{Conclusion}

In this paper, we introduce the H-PRM, an efficient and plug-and-play hotword retriever suitable for both traditional ASR models and Audio LLMs, enabling hotword customization. Experiments show that it outperforms baseline models and handles large-scale hotwords effectively. We also explore prompt effects on Audio LLMs, providing insights and effective prompts. In the future, we will focus on LLM-based contextual ASR and explore more effective methods for hotword customization in Audio LLMs.

\bibliographystyle{unsrtnat}
\bibliography{reference}

\begin{thebibliography}{27}
\providecommand{\natexlab}[1]{#1}
\providecommand{\url}[1]{\texttt{#1}}
\expandafter\ifx\csname urlstyle\endcsname\relax
  \providecommand{\doi}[1]{doi: #1}\else
  \providecommand{\doi}{doi: \begingroup \urlstyle{rm}\Url}\fi

\bibitem[Graves(2012)]{graves2012sequence}
Alex Graves.
\newblock Sequence transduction with recurrent neural networks.
\newblock \emph{arXiv preprint arXiv:1211.3711}, 2012.

\bibitem[Chorowski et~al.(2015)Chorowski, Bahdanau, Serdyuk, Cho, and Bengio]{chorowski2015attention}
Jan~K Chorowski, Dzmitry Bahdanau, Dmitriy Serdyuk, Kyunghyun Cho, and Yoshua Bengio.
\newblock Attention-based models for speech recognition.
\newblock \emph{Computer ence}, 10\penalty0 (4):\penalty0 429--439, 2015.

\bibitem[Graves et~al.(2013)Graves, Mohamed, and Hinton]{graves2013speech}
Alex Graves, Abdel-rahman Mohamed, and Geoffrey Hinton.
\newblock Speech recognition with deep recurrent neural networks.
\newblock In \emph{ICASSP 2013-2013 IEEE International Conference on Acoustics, Speech and Signal Processing (ICASSP)}, pages 6645--6649. IEEE, 2013.

\bibitem[Chan et~al.(2016)Chan, Jaitly, Le, and Vinyals]{chan2016listen}
William Chan, Navdeep Jaitly, Quoc Le, and Oriol Vinyals.
\newblock Listen, attend and spell: A neural network for large vocabulary conversational speech recognition.
\newblock In \emph{ICASSP 2016-2016 IEEE International Conference on Acoustics, Speech and Signal Processing (ICASSP)}, pages 4960--4964. IEEE, 2016.

\bibitem[Waswani et~al.(2017)Waswani, Shazeer, Parmar, Uszkoreit, Jones, Gomez, Kaiser, and Polosukhin]{vaswani2017attention}
A~Waswani, N~Shazeer, N~Parmar, J~Uszkoreit, L~Jones, A~Gomez, L~Kaiser, and I~Polosukhin.
\newblock Attention is all you need.
\newblock \emph{Advances in Neural Information Processing Systems}, 2017.

\bibitem[Gulati et~al.(2020)Gulati, Qin, Chiu, Parmar, Zhang, Yu, Han, Wang, Zhang, Wu, et~al.]{gulati2020conformer}
Anmol Gulati, James Qin, Chung-Cheng Chiu, Niki Parmar, Yu~Zhang, Jiahui Yu, Wei Han, Shibo Wang, Zhengdong Zhang, Yonghui Wu, et~al.
\newblock Conformer: Convolution-augmented transformer for speech recognition.
\newblock \emph{arXiv preprint arXiv:2005.08100}, 2020.

\bibitem[Tian et~al.(2020)Tian, Yi, Tao, Bai, Zhang, and Wen]{tian2020spike}
Zhengkun Tian, Jiangyan Yi, Jianhua Tao, Ye~Bai, Shuai Zhang, and Zhengqi Wen.
\newblock Spike-triggered non-autoregressive transformer for end-to-end speech recognition.
\newblock \emph{arXiv preprint arXiv:2005.07903}, 2020.

\bibitem[Fan et~al.(2021{\natexlab{a}})Fan, Chu, Chang, and Xiao]{fan2021cass}
Ruchao Fan, Wei Chu, Peng Chang, and Jing Xiao.
\newblock Cass-nat: Ctc alignment-based single step non-autoregressive transformer for speech recognition.
\newblock In \emph{ICASSP 2021-2021 IEEE International Conference on Acoustics, Speech and Signal Processing (ICASSP)}, pages 5889--5893. IEEE, 2021{\natexlab{a}}.

\bibitem[Fan et~al.(2021{\natexlab{b}})Fan, Chu, Chang, Xiao, and Alwan]{fan2021improved}
Ruchao Fan, Wei Chu, Peng Chang, Jing Xiao, and Abeer Alwan.
\newblock An improved single step non-autoregressive transformer for automatic speech recognition.
\newblock \emph{arXiv preprint arXiv:2106.09885}, 2021{\natexlab{b}}.

\bibitem[Deng et~al.(2022)Deng, Yang, Watanabe, Higuchi, Cheng, and Zhang]{deng2022improving}
Keqi Deng, Zehui Yang, Shinji Watanabe, Yosuke Higuchi, Gaofeng Cheng, and Pengyuan Zhang.
\newblock Improving non-autoregressive end-to-end speech recognition with pre-trained acoustic and language models.
\newblock In \emph{ICASSP 2022-2022 IEEE International Conference on Acoustics, Speech and Signal Processing (ICASSP)}, pages 8522--8526. IEEE, 2022.

\bibitem[Gao et~al.(2022)Gao, Zhang, McLoughlin, and Yan]{gao2022paraformer}
Zhifu Gao, Shiliang Zhang, Ian McLoughlin, and Zhijie Yan.
\newblock Paraformer: Fast and accurate parallel transformer for non-autoregressive end-to-end speech recognition.
\newblock \emph{arXiv preprint arXiv:2206.08317}, 2022.

\bibitem[Radford et~al.(2023)Radford, Kim, Xu, Brockman, McLeavey, and Sutskever]{radford2023robust}
Alec Radford, Jong~Wook Kim, Tao Xu, Greg Brockman, Christine McLeavey, and Ilya Sutskever.
\newblock Robust speech recognition via large-scale weak supervision.
\newblock In \emph{International conference on machine learning}, pages 28492--28518. PMLR, 2023.

\bibitem[Chu et~al.(2024)Chu, Xu, Yang, Wei, Wei, Guo, Leng, Lv, He, Lin, et~al.]{chu2024qwen2}
Yunfei Chu, Jin Xu, Qian Yang, Haojie Wei, Xipin Wei, Zhifang Guo, Yichong Leng, Yuanjun Lv, Jinzheng He, Junyang Lin, et~al.
\newblock Qwen2-audio technical report.
\newblock \emph{arXiv preprint arXiv:2407.10759}, 2024.

\bibitem[Han et~al.(2021)Han, Dong, Zhou, and Xu]{han2021cif}
Minglun Han, Linhao Dong, Shiyu Zhou, and Bo~Xu.
\newblock Cif-based collaborative decoding for end-to-end contextual speech recognition.
\newblock In \emph{ICASSP 2021-2021 IEEE International Conference on Acoustics, Speech and Signal Processing (ICASSP)}, pages 6528--6532. IEEE, 2021.

\bibitem[Huang et~al.(2023)Huang, Zhang, Yang, Guo, Mu, Xu, and Xie]{huang2023contextualized}
Kaixun Huang, Ao~Zhang, Zhanheng Yang, Pengcheng Guo, Bingshen Mu, Tianyi Xu, and Lei Xie.
\newblock Contextualized end-to-end speech recognition with contextual phrase prediction network.
\newblock \emph{arXiv preprint arXiv:2305.12493}, 2023.

\bibitem[Yang et~al.(2023)Yang, Sun, Wang, Zhang, Ma, and Xie]{yang2023two}
Zhanheng Yang, Sining Sun, Xiong Wang, Yike Zhang, Long Ma, and Lei Xie.
\newblock Two stage contextual word filtering for context bias in unified streaming and non-streaming transducer.
\newblock \emph{arXiv preprint arXiv:2301.06735}, 2023.

\bibitem[Williams et~al.(2018)Williams, Kannan, Aleksic, Rybach, and Sainath]{williams2018contextual}
Ian Williams, Anjuli Kannan, Petar~S Aleksic, David Rybach, and Tara~N Sainath.
\newblock Contextual speech recognition in end-to-end neural network systems using beam search.
\newblock In \emph{Interspeech}, pages 2227--2231. Interspeech, 2018.

\bibitem[Zhao et~al.(2019)Zhao, Sainath, Rybach, Rondon, Bhatia, Li, and Pang]{zhao2019shallow}
Ding Zhao, Tara~N Sainath, David Rybach, Pat Rondon, Deepti Bhatia, Bo~Li, and Ruoming Pang.
\newblock Shallow-fusion end-to-end contextual biasing.
\newblock In \emph{Interspeech}, pages 1418--1422. Interspeech, 2019.

\bibitem[Gourav et~al.(2021)Gourav, Liu, Gandhe, Gu, Lan, Huang, Kalmane, Tiwari, Filimonov, Rastrow, et~al.]{gourav2021personalization}
Aditya Gourav, Linda Liu, Ankur Gandhe, Yile Gu, Guitang Lan, Xiangyang Huang, Shashank Kalmane, Gautam Tiwari, Denis Filimonov, Ariya Rastrow, et~al.
\newblock Personalization strategies for end-to-end speech recognition systems.
\newblock In \emph{ICASSP 2021-2021 IEEE International Conference on Acoustics, Speech and Signal Processing (ICASSP)}, pages 7348--7352. IEEE, 2021.

\bibitem[Han et~al.(2022)Han, Dong, Liang, Cai, Zhou, Ma, and Xu]{han2022improving}
Minglun Han, Linhao Dong, Zhenlin Liang, Meng Cai, Shiyu Zhou, Zejun Ma, and Bo~Xu.
\newblock Improving end-to-end contextual speech recognition with fine-grained contextual knowledge selection.
\newblock In \emph{ICASSP 2022-2022 IEEE International Conference on Acoustics, Speech and Signal Processing (ICASSP)}, pages 8532--8536. IEEE, 2022.

\bibitem[Munkhdalai et~al.(2022)Munkhdalai, Sim, Chandorkar, Gao, Chua, Strohman, and Beaufays]{munkhdalai2022fast}
Tsendsuren Munkhdalai, Khe~Chai Sim, Angad Chandorkar, Fan Gao, Mason Chua, Trevor Strohman, and Fran{\c{c}}oise Beaufays.
\newblock Fast contextual adaptation with neural associative memory for on-device personalized speech recognition.
\newblock In \emph{ICASSP 2022-2022 IEEE International Conference on Acoustics, Speech and Signal Processing (ICASSP)}, pages 6632--6636. IEEE, 2022.

\bibitem[Sainath et~al.(2023)Sainath, Prabhavalkar, Caseiro, Rondon, and Allauzen]{sainath2023improving}
Tara~N Sainath, Rohit Prabhavalkar, Diamantino Caseiro, Pat Rondon, and Cyril Allauzen.
\newblock Improving contextual biasing with text injection.
\newblock In \emph{ICASSP 2023-2023 IEEE International Conference on Acoustics, Speech and Signal Processing (ICASSP)}, pages 1--5. IEEE, 2023.

\bibitem[Pundak et~al.(2018)Pundak, Sainath, Prabhavalkar, Kannan, and Zhao]{pundak2018deep}
Golan Pundak, Tara~N Sainath, Rohit Prabhavalkar, Anjuli Kannan, and Ding Zhao.
\newblock Deep context: end-to-end contextual speech recognition.
\newblock In \emph{2018 IEEE spoken language technology workshop (SLT)}, pages 418--425. IEEE, 2018.

\bibitem[Shi et~al.(2024)Shi, Yang, Li, Chen, Gao, and Zhang]{shi2024seaco}
Xian Shi, Yexin Yang, Zerui Li, Yanni Chen, Zhifu Gao, and Shiliang Zhang.
\newblock Seaco-paraformer: A non-autoregressive asr system with flexible and effective hotword customization ability.
\newblock In \emph{ICASSP 2024-2024 IEEE International Conference on Acoustics, Speech and Signal Processing (ICASSP)}, pages 10346--10350. IEEE, 2024.

\bibitem[Li et~al.(2024)Li, Zhang, Su, Li, Qiao, Ren, Ma, Wei, Tao, and Yang]{li2024multitask}
Yuang Li, Min Zhang, Chang Su, Yinglu Li, Xiaosong Qiao, Mengxin Ren, Miaomiao Ma, Daimeng Wei, Shimin Tao, and Hao Yang.
\newblock A multitask training approach to enhance whisper with open-vocabulary keyword spotting.
\newblock In \emph{Interspeech}, pages 1260--1264. Interspeech, 2024.

\bibitem[Yang et~al.(2024)Yang, Ma, Gao, Zhang, and Chen]{yang2024ctc}
Guanrou Yang, Ziyang Ma, Zhifu Gao, Shiliang Zhang, and Xie Chen.
\newblock Ctc-assisted llm-based contextual asr.
\newblock In \emph{2024 IEEE Spoken Language Technology Workshop (SLT)}, pages 126--131. IEEE, 2024.

\bibitem[Li et~al.(2020)Li, Liu, Wu, Liu, Zhao, and Liu]{li2020robutrans}
Naihan Li, Yanqing Liu, Yu~Wu, Shujie Liu, Sheng Zhao, and Ming Liu.
\newblock Robutrans: A robust transformer-based text-to-speech model.
\newblock In \emph{Proceedings of the AAAI Conference on Artificial Intelligence}, volume~34, pages 8228--8235, 2020.

\end{thebibliography}

\appendix

\end{document}